\documentclass[10pt, journal]{IEEEtran}

\usepackage{graphicx}
\usepackage{amsmath}
\usepackage{booktabs}
\usepackage{makecell}
\usepackage{cite}
\title{Wi-Fi HaLow (IEEE 802.11ah) for Long-Range Monitoring Links:
Point-to-Point NLoS/LoS and LoS Mesh Field Characterization}

\author{Jiajie Xu, Chaabane Mankai, and Mohamed-Slim Alouini}

\begin{document}

\maketitle

%
% Manuscript contents begins.
%

\begin{abstract}
Monitoring deployments often require reliable long-range wireless links to intermittently upload sensor logs and short video snapshots. Wi-Fi HaLow (IEEE~802.11ah) is a promising candidate due to sub-1\,GHz propagation and bandwidth-flexible PHY modes. This summary paper reports a field characterization organized around three deployment-driven regimes: (i) point-to-point Non-Line-of-Sight (NLoS) links; (ii) point-to-point Line-of-Sight (LoS) links over several-hundred-meter distances; and (iii) LoS mesh networking with fixed relay nodes for range extension. Using commodity HaLow \emph{dongle-class} nodes in all regimes, we report application-layer goodput and monitoring-centric update latency based on transferring a representative ``heavy'' object (a $\sim$30\,s video file). The measurements reveal (a) a clear bandwidth--range tradeoff and an NLoS coverage boundary around $\sim$120\,m, (b) gradual throughput decay under LoS up to 814\,m in single-hop with 0.15\,Mbps at the farthest point, and (c) kilometer-class extension under LoS when fixed relays are introduced, reaching 901\,m (two fixed relays) and 1110\,m (three fixed relays).
\end{abstract}

%%%%%%%%%%%%%%%%%%%%%%%%%%%%%%%%%%%%%%%%%%%%%%%%%%%%%%%%%%%%%%%%%%%%%%%%%%%%%%%%%%%%%%%%%%%%%%%%%%%%%%%%%%%%%%%%%%%%%%%%%%%%%%%%%%%%%%%%%%%%%%%%%%%%%%%%
%%%%%%%%%%%%%%%%%%%%%%%%%%%%%%%%%%%%%%%%%%%%%%%%%%%%%%%%%%%%%%%%%%%%%%%%%%%%%%%%%%%%%%%%%%%%%%%%%%%%%%%%%%%%%%%%%%%%%%%%%%%%%%%%%%%%%%%%%%%%%%%%%%%%%%%%
\section{Introduction}
Monitoring-style wireless systems (environmental sensing, infrastructure inspection, perimeter observation, and temporary field operations) often exhibit \emph{bursty} traffic: low-rate telemetry most of the time, punctuated by heavier uploads such as images or short video clips. In these scenarios, the practical requirement is frequently \emph{predictable update time at extended range} rather than peak throughput. For example, a remote camera or inspection node may upload periodic sensor logs and occasionally transmit a short video snippet to confirm events, where completing an update within tens of seconds can be more valuable than achieving very high peak rates only at short range.

\textbf{Why Wi-Fi HaLow (IEEE 802.11ah).}
Wi-Fi HaLow operates in sub-1\,GHz bands and supports bandwidth-flexible PHY modes, aiming to extend coverage while preserving an IP-native Wi-Fi ecosystem \cite{IEEE80211ah,Aust2012,Sun2018,Adame2014}. Compared with conventional Wi-Fi (IEEE 802.11n/ac/ax) that primarily operates at 2.4/5\,GHz, 802.11ah benefits from improved diffraction/penetration and generally lower path loss, which is particularly relevant for outdoor monitoring and cluttered environments where obstacles and foliage dominate propagation. In addition, 802.11ah introduces mechanisms for supporting large numbers of stations and power-saving operation (e.g., restricted access windows and station grouping), aligning with battery-powered monitoring nodes and large-scale sensing \cite{Khorov2015,Adame2014}.

\textbf{Difference from conventional Wi-Fi.}
Conventional Wi-Fi targets high-throughput local connectivity and typically relies on wide channels (20--160\,MHz) at 2.4/5\,GHz. While it can deliver very high data rates, coverage is often limited in outdoor or obstructed environments, and multi-hop extension typically requires additional infrastructure and careful planning. In contrast, 802.11ah trades peak throughput for reach by operating at sub-1\,GHz and allowing narrow channels (e.g., 1/2/4/8\,MHz), making it well matched to ``long-range but still IP-centric'' monitoring links where Mbps-class performance may be available at moderate distances and sub-Mbps operation can remain feasible near the edge \cite{Adame2014,Sun2018}.

\textbf{Difference from LPWA (e.g., LoRa/LoRaWAN).}
LPWA technologies such as LoRa/LoRaWAN are optimized for ultra-long range and ultra-low power but at very low data rates, which fits sparse telemetry yet is inefficient for richer content (images/video snapshots) due to payload and airtime limitations \cite{Raza2017,Augustin2016,Bor2016,Centenaro2016}. Wi-Fi HaLow occupies a middle point: it targets longer range than conventional Wi-Fi while retaining higher throughput and shorter transfer time than LPWA in many practical settings, enabling monitoring workflows that include both sensor logs and occasional ``heavy'' objects over the same link.

\textbf{Target application scenarios.}
This positioning makes 802.11ah attractive for (i) outdoor monitoring (farms, construction sites, solar farms, pipelines); (ii) industrial sensing where sub-1\,GHz penetration helps with obstructions; (iii) temporary deployments (field trials, emergency response) where rapid IP connectivity is desirable; and (iv) range extension via relays when LoS segments can be maintained.

\textbf{What this paper quantifies.}
We organize a compact field characterization into three regimes that map directly to deployment decisions: point-to-point NLoS, point-to-point LoS, and LoS mesh with fixed relays. All three regimes use \emph{dongle-class} HaLow nodes. We report application-layer goodput (Mbps) and monitoring update latency (s) for transferring a representative $\sim$30\,s video file. Concretely, we quantify an NLoS coverage boundary around $\sim$120\,m, single-hop LoS connectivity up to 814\,m, and mesh-enabled LoS extension to 901\,m (two fixed relays) and 1110\,m (three fixed relays).

%%%%%%%%%%%%%%%%%%%%%%%%%%%%%%%%%%%%%%%%%%%%%%%%%%%%%%%%%%%%%%%%%%%%%%%%%%%%%%%%%%%%%%%%%%%%%%%%%%%%%%%%%%%%%%%%%%%%%%%%%%%%%%%%%%%%%%%%%%%%%%%%%%%%%%%%
%%%%%%%%%%%%%%%%%%%%%%%%%%%%%%%%%%%%%%%%%%%%%%%%%%%%%%%%%%%%%%%%%%%%%%%%%%%%%%%%%%%%%%%%%%%%%%%%%%%%%%%%%%%%%%%%%%%%%%%%%%%%%%%%%%%%%%%%%%%%%%%%%%%%%%%%
\section{Point-to-Point NLoS: Bandwidth--Range Tradeoff and Update Latency}
We start with NLoS point-to-point measurements in an obstacle-rich environment, evaluating 1/2/4/8\,MHz to expose the bandwidth--range behavior enabled by IEEE~802.11ah.

\begin{figure}[t]
  \centering
  \includegraphics[width=\linewidth]{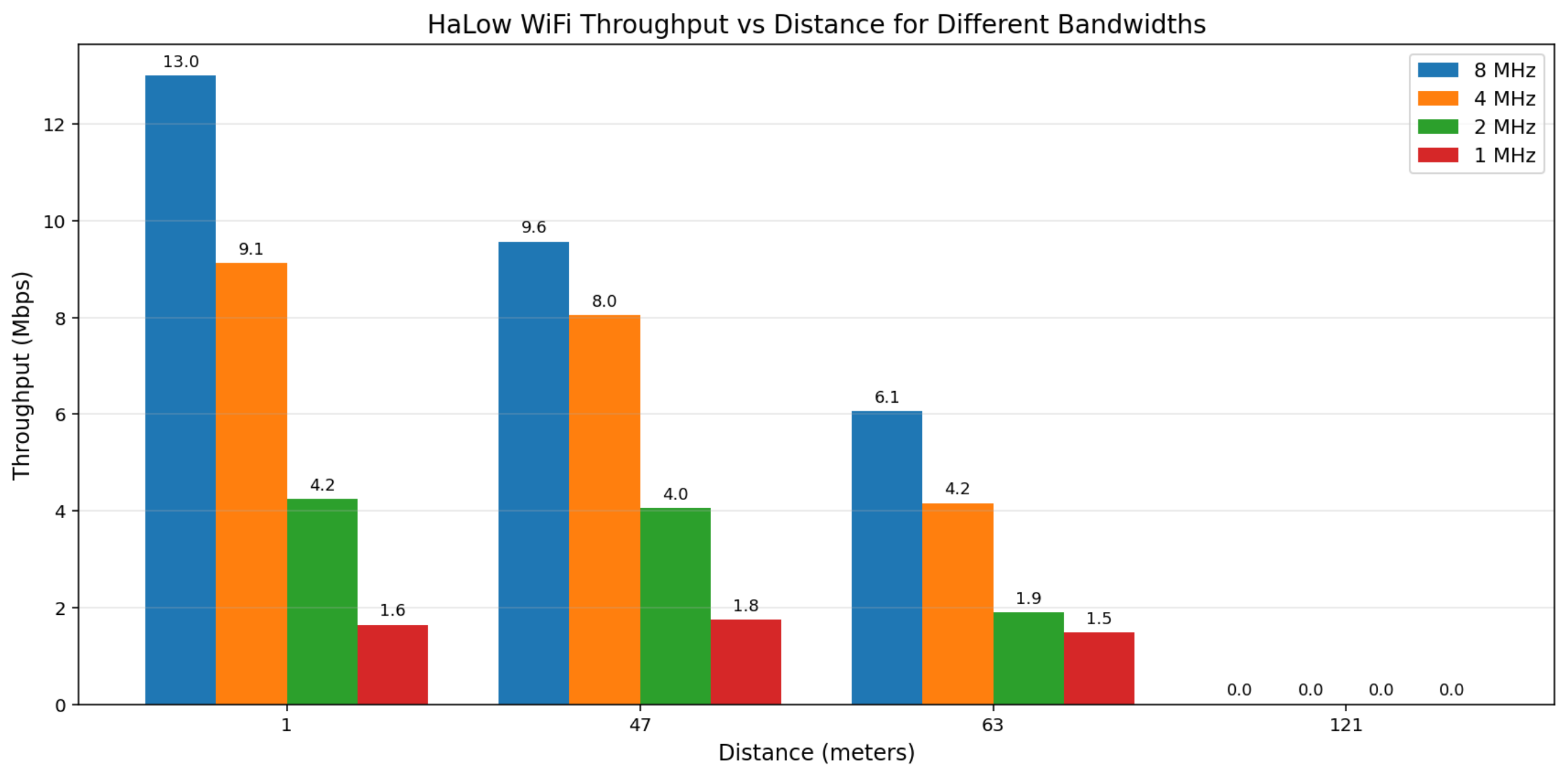}
  \caption{Point-to-point throughput in NLoS for 1/2/4/8\,MHz using dongle-class nodes.}
  \label{fig:fig3}
\end{figure}

Figure~\ref{fig:fig3} shows that wider bandwidth delivers higher short-range throughput, but performance degrades rapidly with distance under NLoS. At 1\,m, throughput is 13.0/9.1/4.2/1.6\,Mbps for 8/4/2/1\,MHz, and drops to 6.1/4.2/1.9/1.5\,Mbps at 63\,m. Beyond a practical boundary (around $\sim$120\,m in this setting), connectivity collapses across bandwidths, indicating that obstacle-induced attenuation dominates and bandwidth narrowing alone cannot sustain operation past the boundary. For monitoring deployments in cluttered sites, this implies that coverage planning and node placement dominate achievable reach, while bandwidth selection is best used to optimize rate \emph{within} the stable operating region.

\begin{figure}[t]
  \centering
  \includegraphics[width=\linewidth]{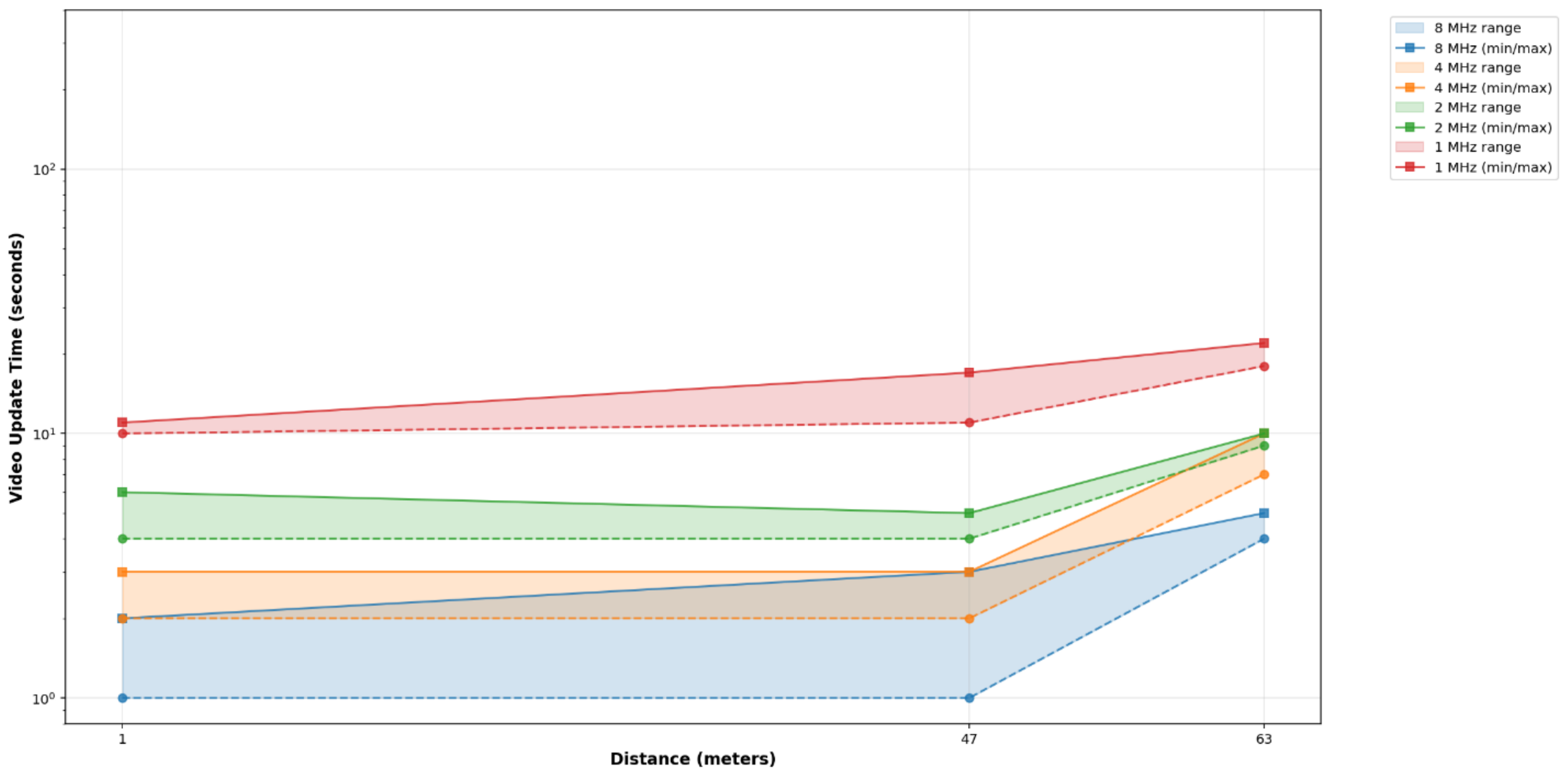}
  \caption{Video update time in NLoS for 1/2/4/8\,MHz using dongle-class nodes (transfer of a representative $\sim$30\,s video file).}
  \label{fig:fig4}
\end{figure}

To translate link behavior into monitoring utility, Figure~\ref{fig:fig4} reports end-to-end transfer time of a representative $\sim$30\,s video file. Within the operational region, transfers typically complete within a few seconds, and the worst-case update time is about 22\,s, which remains within a near real-time window for many monitoring workflows. Near the coverage edge, update time increases sharply and reliability degrades, confirming that the NLoS operating limit is governed primarily by coverage rather than by bitrate tuning alone.

%%%%%%%%%%%%%%%%%%%%%%%%%%%%%%%%%%%%%%%%%%%%%%%%%%%%%%%%%%%%%%%%%%%%%%%%%%%%%%%%%%%%%%%%%%%%%%%%%%%%%%%%%%%%%%%%%%%%%%%%%%%%%%%%%%%%%%%%%%%%%%%%%%%%%%%%
%%%%%%%%%%%%%%%%%%%%%%%%%%%%%%%%%%%%%%%%%%%%%%%%%%%%%%%%%%%%%%%%%%%%%%%%%%%%%%%%%%%%%%%%%%%%%%%%%%%%%%%%%%%%%%%%%%%%%%%%%%%%%%%%%%%%%%%%%%%%%%%%%%%%%%%%
\section{Point-to-Point LoS: Spatial Throughput Distribution up to 814\,m}
We next evaluate LoS point-to-point performance along an outdoor path with increasing distance from the main node to isolate distance effects under favorable propagation.

\begin{figure}[t]
  \centering
  \includegraphics[width=\linewidth]{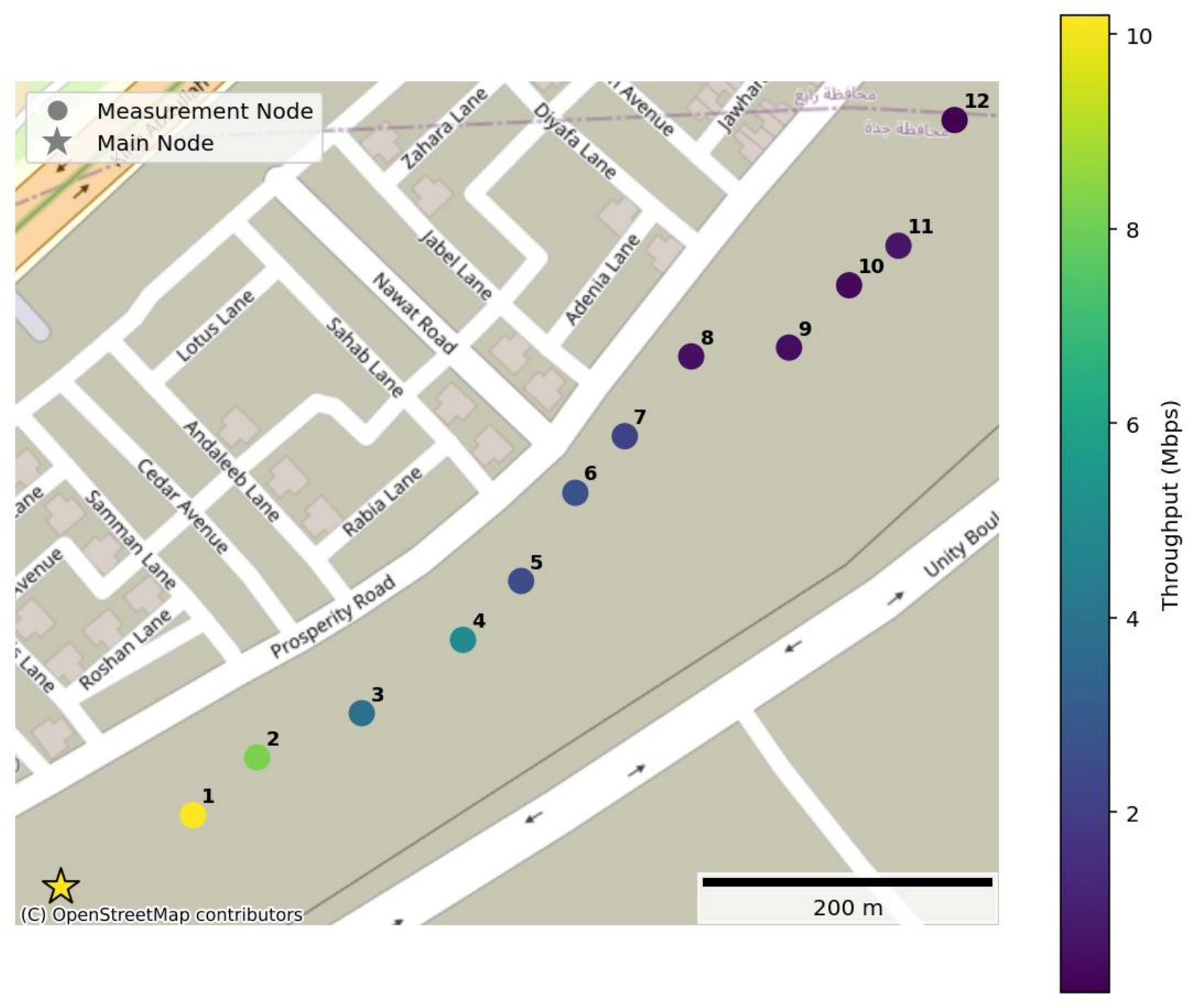}
  \caption{LoS point-to-point throughput map. Star: main node; Circles: measurement locations; Color: throughput (Mbps).}
  \label{fig:fig11}
\end{figure}

Figure~\ref{fig:fig11} exhibits a gradual spatial decay: throughput is highest close to the main node and generally decreases with distance, without the abrupt collapse observed in NLoS. Table~\ref{tab:tab2} reports the measured throughputs at 12 locations. Multi-Mbps performance is sustained in the first few hundred meters (e.g., 9.45\,Mbps at 104\,m; 4.61\,Mbps at 326\,m), and the link remains connected up to 814\,m with 0.15\,Mbps at the farthest point. Although 0.15\,Mbps is low, it still supports periodic low-rate monitoring updates (sensor logs and compressed visual summaries). The non-monotonic points (e.g., 0.73\,Mbps at 729\,m versus 0.38\,Mbps at 685\,m) further indicate that even under LoS, local geometry and multipath can introduce noticeable variability, motivating practical coverage planning based on measured performance.

\begin{table}[t]
\centering
\caption{LoS point-to-point throughput measurements (12 locations).}
\label{tab:tab2}
\small
\setlength{\tabcolsep}{5pt}
\begin{tabular}{c c c}
\toprule
\textbf{Index} & \textbf{Distance (m)} & \textbf{Throughput (Mbps)} \\
\midrule
1  & 104 & 9.45 \\
2  & 163 & 7.63 \\
3  & 240 & 3.55 \\
4  & 326 & 4.61 \\
5  & 382 & 2.28 \\
6  & 448 & 2.51 \\
7  & 499 & 1.99 \\
8  & 569 & 0.53 \\
9  & 626 & 0.47 \\
10 & 685 & 0.38 \\
11 & 729 & 0.73 \\
12 & 814 & 0.15 \\
\bottomrule
\end{tabular}
\end{table}

%%%%%%%%%%%%%%%%%%%%%%%%%%%%%%%%%%%%%%%%%%%%%%%%%%%%%%%%%%%%%%%%%%%%%%%%%%%%%%%%%%%%%%%%%%%%%%%%%%%%%%%%%%%%%%%%%%%%%%%%%%%%%%%%%%%%%%%%%%%%%%%%%%%%%%%%
%%%%%%%%%%%%%%%%%%%%%%%%%%%%%%%%%%%%%%%%%%%%%%%%%%%%%%%%%%%%%%%%%%%%%%%%%%%%%%%%%%%%%%%%%%%%%%%%%%%%%%%%%%%%%%%%%%%%%%%%%%%%%%%%%%%%%%%%%%%%%%%%%%%%%%%%
\section{LoS Mesh Networking: Fixed Relays Extend Coverage Beyond 1\,km}
Single-hop LoS provides several-hundred-meter reach. To push coverage farther without increasing antenna height or transmit power, we introduce fixed relay nodes and evaluate LoS mesh networking. We consider two practical relay densities (two and three fixed relays) and record throughput at selected locations along the path. Multi-hop forwarding is a natural mechanism to extend Wi-Fi coverage when direct links become marginal \cite{IEEE80211s,Akyildiz2005}.

\begin{figure}[t]
  \centering
  \includegraphics[width=\linewidth]{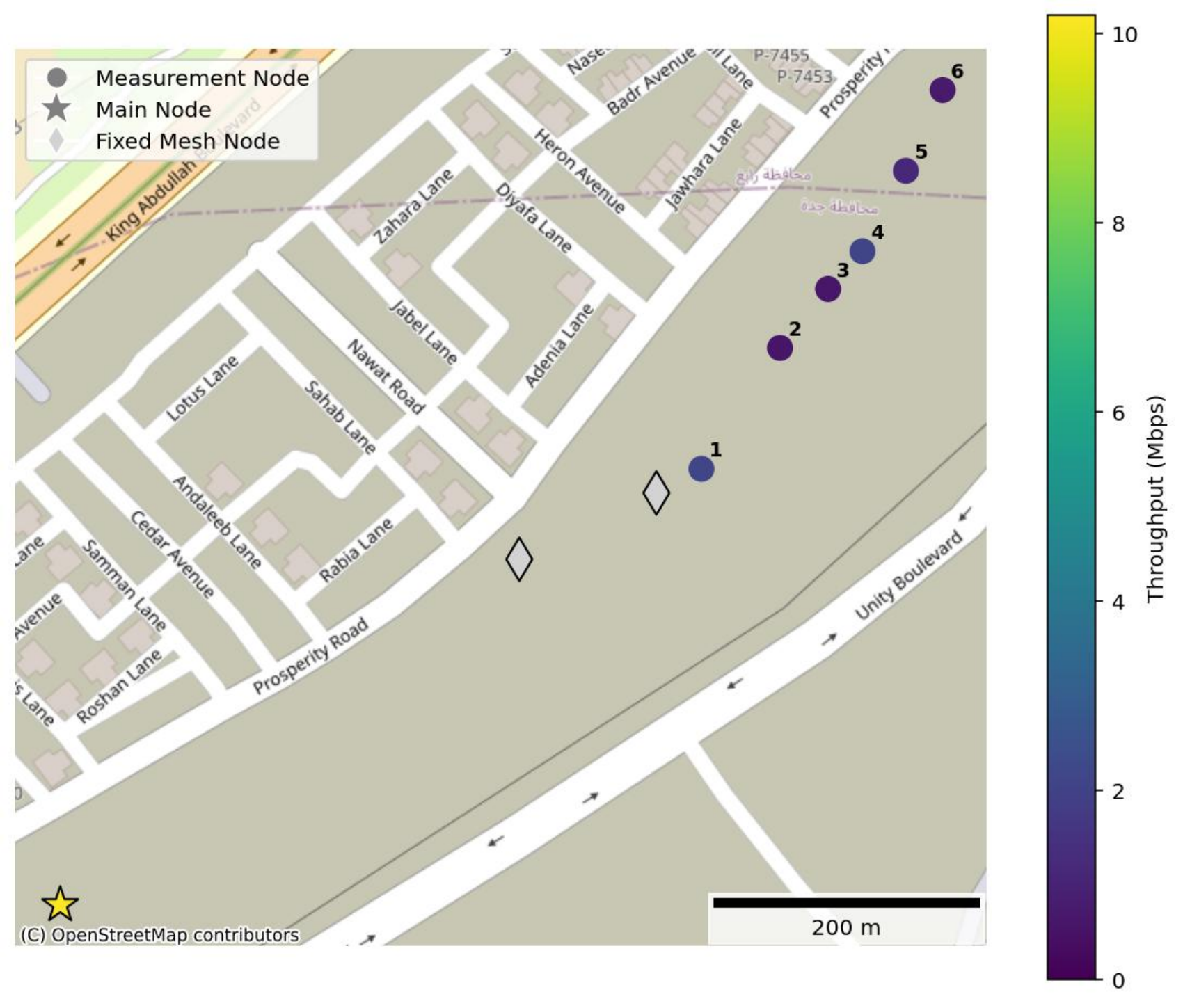}
  \caption{LoS mesh throughput map with two fixed relay nodes. Diamonds: fixed relays; circles: measurement locations; star: main node.}
  \label{fig:fig12}
\end{figure}

With two fixed relays (Figure~\ref{fig:fig12}), the farthest measured point reaches 901\,m with 0.73\,Mbps. Compared with the single-hop LoS farthest point (814\,m at 0.15\,Mbps), relays improve both reach and tail robustness by converting a marginal long hop into multiple stronger LoS hops. Table~\ref{tab:tab4} summarizes relay placement and measured throughputs; notably, several mid/far points remain above 1\,Mbps (e.g., 840\,m at 1.18\,Mbps), indicating that the mesh does not merely ``extend coverage'' but can preserve practical update capability deeper into the edge region when relays are well placed.

\begin{table}[t]
\centering
\caption{LoS mesh results with two fixed relays.}
\label{tab:tab4}
\small
\setlength{\tabcolsep}{5pt}
\begin{tabular}{l c c}
\toprule
\textbf{Node / Index} & \textbf{Distance (m)} & \textbf{Throughput (Mbps)} \\
\midrule
Fixed Node 1 & 431 & -- \\
Fixed Node 2 & 544 & 1.86 \\
\midrule
1 & 582 & 2.13 \\
2 & 683 & 0.56 \\
3 & 739 & 0.62 \\
4 & 776 & 2.09 \\
5 & 840 & 1.18 \\
6 & 901 & 0.73 \\
\bottomrule
\end{tabular}
\end{table}

\begin{figure}[t]
  \centering
  \includegraphics[width=\linewidth]{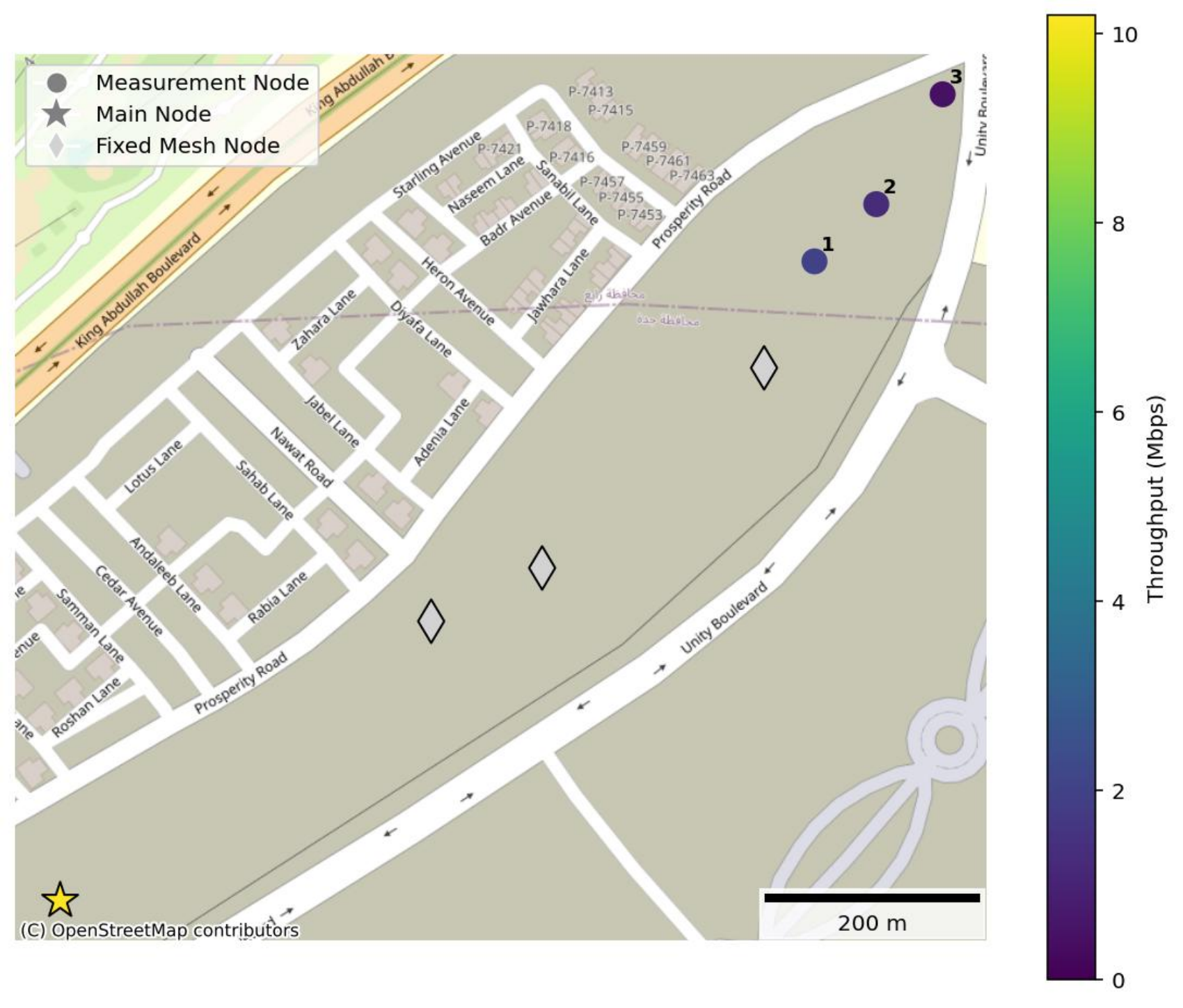}
  \caption{LoS mesh throughput map with three fixed relay nodes, enabling measurements beyond 1\,km.}
  \label{fig:fig13}
\end{figure}

Adding a third fixed relay (Figure~\ref{fig:fig13}) extends the farthest measured distance to 1110\,m with 0.47\,Mbps. Table~\ref{tab:tab5} reports relay placement and the longest-range measurements, showing that beyond-1\,km connectivity remains feasible with sub-Mbps throughput, which is still compatible with periodic monitoring uploads (logs and compressed visual summaries). Overall, the results reinforce a practical guideline: relay placement and density dominate mesh gains; additional nodes provide meaningful extension only when relays preserve strong LoS segments and avoid creating a weak bottleneck hop.

\begin{table}[t]
\centering
\caption{LoS mesh results with three fixed relays.}
\label{tab:tab5}
\small
\setlength{\tabcolsep}{5pt}
\begin{tabular}{l c c}
\toprule
\textbf{Node / Index} & \textbf{Distance (m)} & \textbf{Throughput (Mbps)} \\
\midrule
Fixed Node 1 & 431 & -- \\
Fixed Node 2 & 544 & -- \\
Fixed Node 3 & 820 & 1.59 \\
\midrule
1 & 918  & 2.01 \\
2 & 996  & 1.27 \\
3 & 1110 & 0.47 \\
\bottomrule
\end{tabular}
\end{table}

\textbf{Methodology note.}
Coverage was assessed by recording throughput at selected locations along the path; thus, the spatial maps represent sampled performance rather than a full-area heatmap. Denser sampling would further refine coverage boundaries and quantify directional variability.

%%%%%%%%%%%%%%%%%%%%%%%%%%%%%%%%%%%%%%%%%%%%%%%%%%%%%%%%%%%%%%%%%%%%%%%%%%%%%%%%%%%%%%%%%%%%%%%%%%%%%%%%%%%%%%%%%%%%%%%%%%%%%%%%%%%%%%%%%%%%%%%%%%%%%%%%
%%%%%%%%%%%%%%%%%%%%%%%%%%%%%%%%%%%%%%%%%%%%%%%%%%%%%%%%%%%%%%%%%%%%%%%%%%%%%%%%%%%%%%%%%%%%%%%%%%%%%%%%%%%%%%%%%%%%%%%%%%%%%%%%%%%%%%%%%%%%%%%%%%%%%%%%

\section{Conclusion and Future Work}
This paper presented a field characterization of Wi-Fi HaLow for monitoring links across three deployment-driven regimes using dongle-class nodes throughout. In NLoS, we observed a pronounced coverage boundary around $\sim$120\,m and a bandwidth--range tradeoff that primarily benefits operation within the stable region; monitoring-oriented video updates complete within a few seconds in-range, with a worst case around 22\,s. In LoS, throughput decays gradually and remains connected up to 814\,m (0.15\,Mbps at the farthest point). Introducing fixed relays enables kilometer-class extension under LoS, reaching 901\,m with two relays (0.73\,Mbps) and 1110\,m with three relays (0.47\,Mbps), underscoring relay placement and density as the dominant design variables for mesh-enabled range extension.

In the future, we would plan to (i) perform denser and multi-directional sampling to turn the current path-based measurements into area coverage statistics (including variability over time and different site geometries); (ii) jointly measure throughput, packet loss, and monitoring-object completion time under controlled traffic patterns (telemetry-only, mixed telemetry  and bursty video); (iii) incorporate energy profiling under duty-cycled operation to connect link performance with battery lifetime, (iv) study relay placement and routing/forwarding policies that maximize tail throughput and update reliability (e.g., bottleneck-aware placement), and (v) provide a systematic cross-technology comparison against conventional 2.4\,GHz Wi-Fi solutions and LPWA baselines (e.g., LoRa/LoRaWAN) for representative monitoring workloads.

%%%%%%%%%%%%%%%%%%%%%%%%%%%%%%%%%%%%%%%%%%%%%%%%%%%%%%%%%%%%%%%%%%%%%%%%%%%%%%%%%%%%%%%%%%%%%%%%%%%%%%%%%%%%%%%%%%%%%%%%%%%%%%%%%%%%%%%%%%%%%%%%%%%%%%%%
%%%%%%%%%%%%%%%%%%%%%%%%%%%%%%%%%%%%%%%%%%%%%%%%%%%%%%%%%%%%%%%%%%%%%%%%%%%%%%%%%%%%%%%%%%%%%%%%%%%%%%%%%%%%%%%%%%%%%%%%%%%%%%%%%%%%%%%%%%%%%%%%%%%%%%%%

%%%%%%%%%%%%%%%%%%%%%%%%%%%%%%%%%%%%%%%%%%%%%%%%%%%%%%%%%%%%%%%%%%%%%%%%%%%%%%%%%%%%%%%%%%%%%%%%%%%%%%%%%%%%%%%%%%%%%%%%%%%%%%%%%%%%%%%%%%%%%%%%%%%%%%%%
%%%%%%%%%%%%%%%%%%%%%%%%%%%%%%%%%%%%%%%%%%%%%%%%%%%%%%%%%%%%%%%%%%%%%%%%%%%%%%%%%%%%%%%%%%%%%%%%%%%%%%%%%%%%%%%%%%%%%%%%%%%%%%%%%%%%%%%%%%%%%%%%%%%%%%%%
%
% Note that the authors' affiliations and contact information should be
% here, after the references.
%
\noindent\small
Jiajie Xu, Chaabane Mankai, and Mohamed-Slim Alouini are with the Computer, Electrical, and Mathematical Sciences and Engineering Division (CEMSE), King Abdullah University of Science and Technology (KAUST), Thuwal 23955, Saudi Arabia; e-mail: jiajie.xu.1@kaust.edu.sa; chaabane.mankai@kaust.edu.sa;  slim.alouini@kaust.edu.sa.


\begin{thebibliography}{9}
  
\bibitem{IEEE80211ah}
IEEE Standard Association, ``IEEE Standard for Information Technology---Telecommunications and Information Exchange Between Systems Local and Metropolitan Area Networks---Specific Requirements Part 11: Wireless LAN Medium Access Control (MAC) and Physical Layer (PHY) Specifications, Amendment 2: Sub 1~GHz License Exempt Operation,'' \emph{IEEE Std 802.11ah-2016}, 2016.

\bibitem{Adame2014}
T.~Adame, A.~Bel, A.~Carreras, J.~L.~Meli\`a-Segu\'\i, M.~Oliver, and R.~Pous, ``IEEE 802.11ah: The WiFi Approach for M2M Communications,'' \emph{IEEE Wireless Communications}, vol.~21, no.~6, pp.~144--152, 2014.

\bibitem{Aust2012}
S.~Aust, R.~V.~Prasad, and I.~G.~M.~M.~Niemegeers, ``IEEE 802.11ah: The Wi-Fi for the Internet of Things,'' in \emph{Proc. IEEE ICC Workshops}, 2012.

\bibitem{Sun2018}
W.~Sun, M.~Choi, and S.~Choi, ``IEEE 802.11ah: A Long Range 802.11 WLAN at Sub 1~GHz,'' \emph{ICT Express}, vol.~4, no.~2, pp.~66--73, 2018.

\bibitem{Khorov2015}
E.~Khorov, A.~Krotov, and A.~Lyakhov, ``A Survey on IEEE 802.11ah: An Enabling Networking Technology for Smart Cities,'' \emph{Computer Communications}, vol.~58, pp.~53--69, 2015.

\bibitem{Raza2017}
U.~Raza, P.~Kulkarni, and M.~Sooriyabandara, ``Low Power Wide Area Networks: An Overview,'' \emph{IEEE Commun. Surveys \& Tutorials}, vol.~19, no.~2, pp.~855--873, 2017.

\bibitem{Centenaro2016}
M.~Centenaro, L.~Vangelista, A.~Zanella, and M.~Zorzi, ``Long-Range Communications in Unlicensed Bands: The Rising Stars in the IoT and Smart City Scenarios,'' \emph{IEEE Wireless Communications}, vol.~23, no.~5, pp.~60--67, 2016.

\bibitem{Augustin2016}
A.~Augustin, J.~Yi, T.~Clausen, and W.~Townsend, ``A Study of LoRa: Long Range \& Low Power Networks for the Internet of Things,'' \emph{Sensors}, vol.~16, no.~9, 2016.

\bibitem{Bor2016}
M.~Bor, U.~Roedig, T.~Voigt, and J.~M.~Alonso, ``Do LoRa Low-Power Wide-Area Networks Scale?,'' in \emph{Proc. ACM MSWiM}, 2016.

\bibitem{IEEE80211s}
IEEE Standard Association, ``IEEE Standard for Information Technology---Telecommunications and Information Exchange Between Systems Local and Metropolitan Area Networks---Specific Requirements Part 11: Wireless LAN Medium Access Control (MAC) and Physical Layer (PHY) Specifications, Amendment: Mesh Networking,'' \emph{IEEE Std 802.11s-2011}, 2011.

\bibitem{Akyildiz2005}
I.~F.~Akyildiz, X.~Wang, and W.~Wang, ``Wireless Mesh Networks: A Survey,'' \emph{Computer Networks}, vol.~47, no.~4, pp.~445--487, 2005.

\end{thebibliography}
\end{document}